\documentstyle[twocolumn]{article} 

\def\kms{km s$^{-1}$}
\def\Msun{M_\odot}
\def\mpc{M_\odot{\rm pc}^{-2}}

\begin{document} 

\title{Dark Bulge, Exponential Disk, and Massive Halo 
in the Large Magellanic Cloud}

\author{
Yoshiaki SOFUE \\
Inst. of Astronomy, Univ. of Tokyo, Mitaka, Tokyo 181-8588 \\
E-mail: sofue@ioa.s.u-tokyo.ac.jp \\
\\
(1999 PASJ in press)
}

\date{}
\maketitle

\begin{abstract}
The rotation curve of the Large Magellanic Cloud, which we have 
derived from high-resolution HI position-velocity diagrams
observed by Kim et al (1998), shows a steep central rise and flat rotation
with a gradual rise toward the edge.
Using the rotation curve, we have calculated the distribution of 
surface mass density, and show that the LMC has a dark compact bulge, 
an exponential disk, and a massive halo. 
The bulge is 1.2 kpc away from the center of the stellar bar, and is not
associated with an optical counterpart.
This indicates that the "dark bulge" has a large fraction of dark matter,
with an anomalously high mass-to-luminosity (M/L) ratio.
On the contrary, the stellar bar has a smaller M/L ratio compared to the
surrounding regions.
\vskip 2mm

\end{abstract}

\section{Introduction}

Rotation curves of dwarf galaxies are known to increase
monotonically toward their outer edge, and the mass-to-luminosity ratio,
and therefore, the dark matter fraction
is larger than that in usual spiral galaxies (Persic et al 1996).
However, little is known of their internal distribution of dynamical mass, 
because of the insufficient spatial resolution due to their 
small angular extents as well as for the low surface brightness in HI and
 H$\alpha$ lines.

HI observations of the Large Magellanic Cloud (LMC), the nearest dwarf 
galaxy, have provided us with a unique opportunity 
to study detailed kinematics (e.g., Luks and Rohlfs 1992; Putman et al 1998).
Recently, Kim et al (1998) have obtained high-resolution HI kinematics 
from interferometer observations.
They have shown that the rotation characteristics is nearly axisymmetric,
and this fact allows us to derive a reliable rotation curve by using a 
position-velocity diagram along the major axis.

In the present paper, we derive a new rotation curve for the LMC, and  
discuss the distribution of dynamical mass and dark matter. 

\section{Position-Velocity Diagrams and Rotation Curve}

Fig. 1 shows the HI velocity field as reproduced from Kim et al (1998),
superposed on a smoothed DSS (Digitized Sky Survey) image in B band.
The cross indicates the kinematical center defined by Kim et al..
We have, here, determined a position, 
RA=5h 13.8m, Dec=$-68^\circ 38'$ (J2000), as indicated by the asterisk,
at which the velocity gradient attains the maximum, and around 
which the velocity field becomes most symmetric. 
From the symmetric velocity field, we may assume that the gas
is circularly rotating around this  kinematical center in the central 2 kpc
radius region.
The interferometer observations may miss the intensities from extended
gas disk. 
However, since the extended gas and the gaseous features 
detected with the interferometer are thought to be rotating at the 
same speed, we assume that the observed velocity field represents
that of the total gas.

In Fig. 2 we reproduce the PV diagram across a position close to the
adopted kinematical center from Kim et al (1998).
The thin line represents a rotation curve fitted by Kim et al. by using
an AIPS utility, which adopts a simple function. 
However, the innermost part of the PV diagram shows a much steeper variation 
from positive to negative velocities, which appears to be 
not well traced.
In order to obtain a more precise rotation velocities in the innermost region, 
we apply the envelope-tracing method of maximum-velocities (Sofue 1996),
which we have done by eye-estimates on the PV diagram. 
The newly determined rotation velocities are indicated by the thick curve,
by which the steep velocity variation is now well traced.
Besides the bright features traced by this curve, a higher-velocity feature
is found at -60 to -70 \kms\ and $+20'$.
If we trace this feature, the velocity variation
around the center becomes steeper, and the rotation curve will have 
much steeper rise at the center.

Using the thick curve in Fig. 2 and correcting for the inclination angle 
of 33$^\circ$, we obtain a rotation curve as shown in Fig. 3.
The outer rotation curves beyond the HI disk up to 8 kpc
has been taken from Kunkel et al (1997).
In the following discussion, we adopt the thick curve in Fig. 3,
which is drawn by smoothing the observed curve (thin line).
The thus-obtained rotation curve is found to be similar to those of 
usual disk galaxies (Rubin et al 1982; Sofue 1996, 1997; Sofue et al
1998, 1999) except for the absolute values.
The rotation curve of the LMC is characterized by

{\parindent=0pt
(1) a steep central rise, 

(2) a flat part in the disk, and 

(3) a gradual rise toward the edge.
}

\section{Surface Mass Distribution: Bulge, Disk, and Massive Halo}

Given a rotation curve from the center to the outer edge
with a sufficient resolution, we can directly calculate 
the surface mass density (SMD)(Takamiya and Sofue 1999).
This method is not intervened by any models such as the Plummer
potentials and exponential disks. 
An extreme case is to assume a spherical symmetry:
the rotation velocity is used to calculate the total mass involved 
within a radius, which is then used to calculate the SMD.
Another extreme case is to assume a thin rotating disk:
the SMD can be directly calculated by using 
the Poisson equation (Binney and Tremaine 1987).
We assume that the true mass distribution will lie in between these two cases,
which are indeed found to coincide within a factor of 1.5.  
Fig. 4 shows the calculated SMD for the disk (full line) 
and spherical cases (dashed line).
The full line will better represent a disk component as the HI appearance
indicates, while the spherical case would be better for an inner bulge.
The obtained SMD distribution can be summarized as the following.

\noindent{\it (1) Dark bulge}

The distribution of SMD shows a compact, high-density component
tightly concentrated around the kinematical center. 
This component is not associated with an optical counterpart, and
we call it a "dark bulge".
The $e$-folding scale radius is about 120 pc, when fitted with
an exponential function.
The mass involved within 240 pc radius, at which the SMD
becomes equal to that of the disk component, is
$\sim 1.7\times10^8M_\odot$.
Except for its small radius, the dark bulge is as dense as a central 
bulge of normal disk galaxies:
the SMD is estimated to be
$\sim 1.0\times10^3 \mpc$ 
at its scale radius, which is comparable to or only 
by a factor of ten smaller than the values of usual galactic 
bulges ($\sim 10^{3-4} \mpc$) at their scale radii (a few hundred pc) 
(Takamiya and Sofue 1999).

\noindent{\it (2) Exponential disk}

The SMD, then, is followed by an exponentially decreasing part with an 
$e$-folding scale radius of 1.0 kpc.
This "exponential disk" in mass is dominant up to a radius of 2 kpc,
and extends to 3 to 4 kpc radius, being gradually replaced
by an extended halo component.
The total mass involved within 2 kpc radius is $\sim 2\times10^9M_\odot$, 
consistent with the value obtained by Kim et al (1998).
The SMD of $\sim 200 \Msun{\rm pc}^{-2}$
at its scale radius is comparable to the values of usual disk galaxies
($\sim 300 - 500 \mpc$)  at their scale radii ($\sim 3-5$ kpc).

The total $B$ magnitude of LMC is 0.6 mag. within an angular extent of  
$650' \times 550'$, or within $\sim 4.3$ kpc radius 
(de Vaucouleurs et al 1991).
For a distance modulus of 18.50 (50 kpc) and $B-V$ color of 0.55, 
we obtain a  $V$ magnitude $-18.45$.
This leads to a $V$ band luminosity of $\sim 2 \times 10^9 L_\odot$.
Since the mass involved within 4.3 kpc is $\sim 4.5 \times 10^9 \Msun$, we 
obtain a mean M/L ratio of about 2.2 for the disk component. 

\noindent{\it (3) Massive halo}

The SMD in the outer halo obeys an exponentially
decreasing function with an $e$-folding scale radius of 5.2 kpc.
The total mass involved within 8 kpc radius is $\sim 1\times10^{10}M_\odot$. 
Since the outer envelope is not well visible in optical photographs (DSS), 
this may suggest again a high M/L ratio in the halo.
This is consistent with the high M/L ratio in many other dwarf galaxies,
for which gradually-rising rotation curves are generally observed 
(Persic et al 1996). 
Kunkel et al (1997) have suggested that the high velocities in the outer 
region could be due to a tidal effect with the Small Magellanic Cloud. 
Alternatively, tidal warping could also cause apparently higher (or lower) 
velocities, because the galaxy is nearly face on.
In these cases, the presently estimated  mass would be significantly  
overestimated.

\section{Discussion}

The dark bulge is significantly displaced from the stellar bar.
No optical counterpart is visible, despite that the extinction is supposed to
be not large for the nearly face-on orientation.
Since the surface luminosity appears to be not significantly 
enhanced from the average in the disk part, where M/L ratio is $\sim 2$, 
the bulge's M/L would be as large as $ \sim 20 - 50 M_\odot/L_\odot$.
Such a high M/L indicates a significant excess of the dark matter fraction
in the bulge. 
As discussed in Section 2,  if the  faint and higher-velocity features 
in the PV diagrams are taken into account, the rotation curve will have a much 
steeper rise in the center than that traced by the thick line in Fig. 2.
If this is the case, the bulge's mass density will be still higher, and so is 
the dark mass fraction.

The center of the stellar bar is about 1.2 kpc away from the center
of the dark bulge (Fig. 1). 
No significant mass enhancement, which might be related to the bar, is 
detected at radii 1 to 2 kpc in Fig. 4. 
In fact, neither a stream motion in the velocity field, 
nor an anomaly in the PV diagrams are observed along the 
stellar bar  (Kim et al 1998).
We may argue that the M/L ratio in the stellar bar is 
significantly lower than that in the surrounding regions. 

We finally comment on an alternative possibility to explain the 
displacement of the optical and  kinematical centers.
Ram-pressure stripping of the gas disk either due to an intergalactic wind
(Sofue 1994) and/or a gaseous inflow from the Small Magellanic Cloud 
(Gardiner et al 1994; Putman et al 1998) could cause such a displacement.  
In order for the HI velocity field being kept unperturbed, 
it must occur as quickly as in a crossing time of the innermost disk, or 
within $ t \sim  r / v_{\rm rot} \sim 2 \times 10^6$ years, 
where $r\sim 120$ pc is the radius of the innermost HI disk
and $v_{\rm rot}\sim 55$ \kms is the rotating velocity. 
This requires a  wind velocity greater than 
$v_{\rm ram} \sim d /t \sim 500$ \kms, where $d\sim 1$ kpc is the displacement.
Such a high velocity would be possible for an intergalactic wind. 
However, the ram-stripping condition,
  $\rho_{\rm ram} v_{\rm ram}^2 > \rho_{\rm HI} v_{\rm rot}^2$,
seems to be not satisfied: 
for $\rho_{\rm ram} \sim 10^{-4}m_{\rm H}$ cm$^{-3}$, 
$v_{\rm ram}\sim 500$ \kms,
$\rho_{\rm HI} \sim 0.1 m_{\rm H}$ cm$^{-3}$, and 
$v_{\rm rot} \sim 55$ \kms, 
we have $0.5\times 10^{-12}$ and $5\times 10^{-12}$ dyne cm$^{-2}$
for the first and second terms, respectively. 
Therefore, the ram-stripping appears not a likely origin.
In any other circumstances, where no gravitational attraction is present,
the inner rotation feature will disappear within about one million years.
So, in so far as the velocity field is assumed to represent rotation, 
a high-density mass concentration at the kinematical center, and therefore,
the presence of a dark bulge, will be an inevitable consequence. 

Another possible idea is to attribute the velocity gradients in the
kinematical center to local velocity anomalies such as due to wiggles
by HI shells and/or turbulence.
However, Kim et al's modulus map of velocity residuals indicates that 
density waves and warping are more dominant to cause the residuals. 
Neither the largest HI shells nor 30 Dor are 
associated with the largest velocity residuals and wiggles.
Moreover, there appear no particular HI shells or associated
starforming regions around the kinematical center where the velocity
variation is steepest.

Hence, we here rely more on our assumption that the HI kinematics of the
LMC directly manifests the dynamics of the galaxy.
We also comment that, if we adopt the smooth rotation curve of Kim et al. 
(1998) as an alternative case, the mass distribution would 
be similar to the one represented by the disk component in Fig. 4.
Even in this case the mass center is still significantly displaced from the
optical bar, and no particular optical enhancement corresponding to the
exponential disk is recognized around the kinematical center.

\vskip 2mm
{The author thanks T. Takamiya for calculating the SMD from
the rotation curve data. He is also indebted to Prof. M.
Fujimoto of Nagoya University for invaluable discussion.}

\section*{References} 

\def\r{\hangindent=1pc  \noindent}

\r Binney, T., Tremaine, S. 1987, in Galactic Astronomy (Princeton
	Univ. Press).

\r de Vaucouleurs G., de Vaucouleurs A., Corwin  H. G. Jr., et al.,
   1991, in {\it  Third Reference Catalogue of Bright Galaxies}
   (New York: Springer Verlag)

\r Gardiner, L. T., Sawa, T.  Fujimoto, M. 1994 NMRAS, 266, 567.  

\r Kim, S., Stavely-Smith, L., Dopita, M. A., Freeman, K. C.,
	Sault, R. J., Kesteven, M. J., and McConnell, D. 1998, ApJ 503, 674.

\r Kunkel, W. E., Demers, S., Irwin, M. J., and Albert, L. 1997, ApJ 488, L129. 

\r Luks, Th. and Rohlfs, K. 1982 AA, 263, 41L. @

\r Persic, M., Salucci, P., Stel, F.  1996, MNRAS, 281,  27. 

\r Putman, M. E., Gibson, B. K., Staveley-Smith, L., Banks, C., Barnes, D. G.
 	et al. 1998 Nature 394, 752.    

\r Rubin, V. C., Ford, W. K., Thonnard, N. 1982, ApJ, 261, 439

\r Sofue, Y. 1994, PASJ, 46, 431 

\r Sofue, Y. 1996, ApJ, 458, 120

\r Sofue, Y. 1997, PASJ, 49, 17

\r Sofue, Y., Tomita, A.,  Honma, M.,Tutui, Y. and Takeda, Y.
        1998, PASJ 50, 427. 

\r Sofue, Y., Tutui, Honma, M., Tomita, A., Takamiya, T., Koda, J. and 
	and Takeda, Y.  1999, ApJ. Sept. issue. 

\r Takamiya, T., and Sofue, Y. 1999, submitted to ApJ.

\label{last}

\vskip 10mm
\parindent=0pt
\parskip=5mm

Figure Captions
 
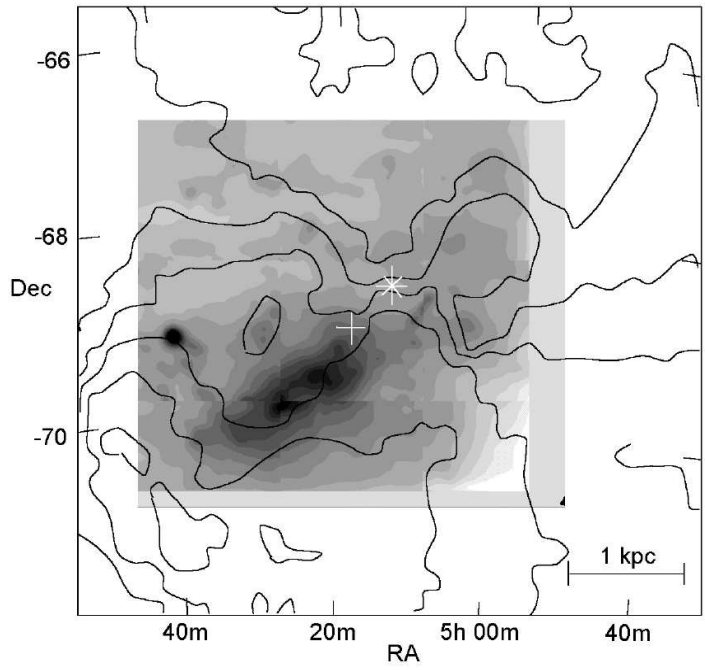
Fig. 1: HI velocity field of LMC (Kim et al 1998)
superposed on a smoothed DSS image in B band.
The cross indicates the kinematical center of Kim et al (1998), and the
asterisk is the adopted center position in this paper.
Note the significant displacement of the bar from the kinematical center.

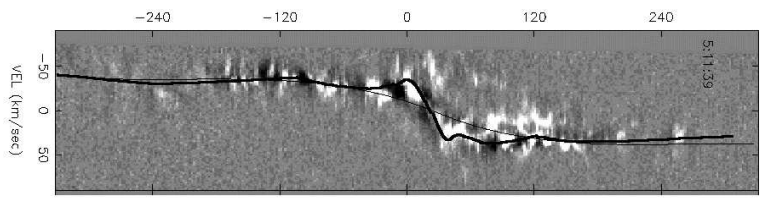
Fig.2: Position-velocity diagram of the LMC along the major axis 
(Kim et al 1998).
We trace the maximum-velocity envelope by the thick line to derive the 
rotation curve.

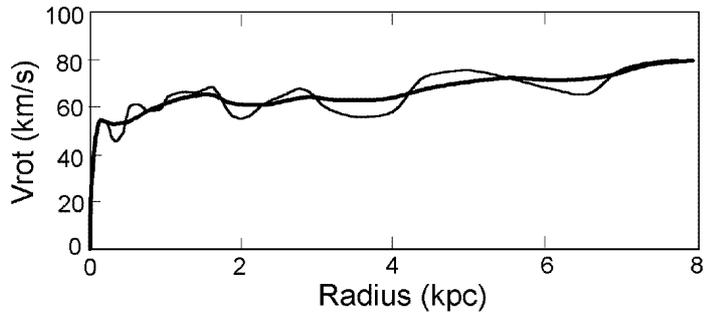
Fig.3: Rotation curve of the LMC. Note the steep central rise, and
flat disk rotation. The thin line denotes the observation, 
and the thick line is a smoothed rotation curve, which we adopt 
for calculating the surface-mass distribution. 

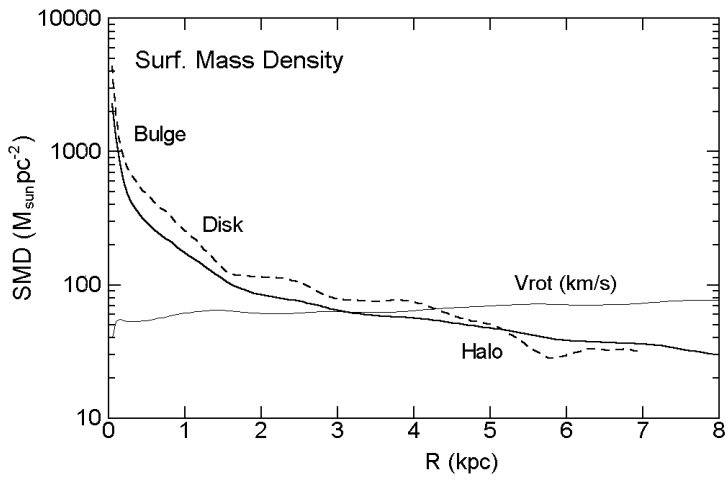
Fig.4: Radial distributions of the surface mass density for a thin disk
assumption (full line) and for a spherical assumption (dashed line).  
There appear three components: 
(1) a dense and compact bulge at radii $<240$ pc, 
(2) an exponential disk until 2 to 3 kpc, and 
(3) a massive halo.

\end{document}